\documentstyle[12pt]{ioplppt}
\input epsf.tex

\begin{document}
\jl{1}

\title{The McCoy-Wu Model in the Mean-field Approximation}

\author{Bertrand Berche\dag, Pierre Emmanuel Berche\dag\ddag, 
Ferenc Igl\'oi\S$\parallel$\ and G\'abor Pal\'agyi\P}
\address{\dag\ Laboratoire de Physique des Mat\'eriaux\ftnote{6}{Unit\'e de
Mixte de Recherche  CNRS No 7556}, Universit\'e Henri
Poincar\'e, Nancy 1\\
B.P. 239, F-54506 Vand\oe uvre les Nancy, France}
\address{\ddag\ Institut f\"ur Physik, Johannes Gutenberg-Universit\"at, Mainz
Staudinger Weg 7,\\ 55099 Mainz, Germany}
\address{\S\ Research Institute for Solid State Physics, 
H-1525 Budapest, P.O.Box 49, Hungary}
\address{$\parallel$ Institute for Theoretical Physics,
Szeged University, H-6720 Szeged, Hungary}
\address{\P\ Department of Physics,
University of Veszpr\'em, H-8201 Veszpr\'em, Hungary}

\date{\today}

%\date{January, 1998}

\begin{abstract}
\baselineskip=16pt
We consider a system with randomly layered ferromagnetic bonds
(McCoy-Wu model) and study its critical properties in the frame of mean-field theory. In the
low-temperature phase there is an average spontaneous magnetization in the
system, which vanishes as a power law at the critical point with the
critical exponents $\beta \approx 3.6$ and $\beta_1 \approx 4.1$ in
the bulk and at the surface of the system, respectively. The singularity
of the specific heat is characterized by an exponent $\alpha \approx -3.1$.
The samples reduced critical temperature $t_c=T_c^{av}-T_c$ has a power law
distribution $P(t_c) \sim t_c^{\omega}$ and we show that the difference
between the
values of the critical exponents in the pure and in the random system is just
$\omega \approx 3.1$. Above the critical temperature the thermodynamic
quantities behave analytically, thus the system does not exhibit Griffiths
singularities.
\end{abstract}

\newcommand{\bc}{\begin{center}}
\newcommand{\ec}{\end{center}}
\newcommand{\be}{\begin{equation}}
\newcommand{\ee}{\end{equation}}
\newcommand{\beqn}{\begin{eqnarray}}
\newcommand{\eeqn}{\end{eqnarray}}

\maketitle

\section{Introduction}
\label{sec:intro}

More than a quarter century ago McCoy and Wu~\cite{mccoywu} have introduced
and partially solved a randomly layered Ising model on the square lattice.
In the model, the nearest-neighbour vertical couplings $K$ are the same,
whereas the horizontal couplings $J_i$ are identical within each column,
but vary from column to column, such that they are taken independently
from a distribution $\rho(J)\d J$. Recently, the solution of the McCoy-Wu (MW)
model and the related random transverse-field Ising spin chain have been
substantially extended by renormalization group~\cite{fisher} and
numerical~\cite{youngrieger,igloirieger97,young,igloirieger98} studies.
Exact values for the average bulk $\beta$ and surface $\beta_1$ magnetization
exponents and the $\nu$ correlation length exponent are given by:
\be
\beta={3-\sqrt{5} \over 2},~~~\beta_1=1~~~{\rm and}~~~\nu=2\;,
\label{exponents}
\ee
which all differ from the corresponding values in the pure system.
We note that several physical quantities of the MW model are not self-averaging
at the critical point, consequently their typical and average values are
different. Further curiosity of the MW model is the existence of Griffiths-McCoy
singularities~\cite{griffiths,mccoy} at both sides of the critical point, where the
vertical spin-spin correlations decay as a power law with temperature-dependent decay exponents and, consequently, the susceptibility is divergent
in a whole region.

The MW model, more precisely its quantum version, has been generalized for
higher dimensions; namely quantum spin glasses in 2 and 3 space dimensions~\cite{qspglass}, the corresponding mean-field 
theory~\cite{MFSG}, diluted
transverse Ising ferromagnets in higher dimensions~\cite{DFM} and random
bond Ising ferromagnets in $d=2$~\cite{RFM}. In all of these models, disorder
is uncorrelated in the $d$ space dimensions and perfectly correlated in the
additional imaginary time direction. Various analytical techniques, known
from classical spin glasses~\cite{spinglass}, are at hand to treat the
mean-field theory of other cases~\cite{MFSG}.
 
In this paper we consider a different type of generalization of the
MW model
to $d>2$ dimensions. 
In our approach the variation in the $J_i$ couplings
remains one-dimensional and these couplings are identical in
$(d-1)$-dimensional columns, while couplings in the other
$(d-1)$-directions are the same, $K$. We study the problem within mean-field
theory, therefore we call our system as Mean-Field McCoy-Wu (MFMW) model.
We mention that inhomogeneous layered systems with quasi-periodic and smoothly varying
interactions have been recently studied in the frame of mean-field theory by similar 
methods~\cite{inhomogen1,inhomogen2}.

The paper 
is organized as
follows: In \sref{sec:sec2}, we present the model and the numerical technique which
is used to obtain the order parameter profile. The critical exponents are
determined in 
\sref{sec:sec3}, while in \sref{sec:sec4} an analysis of
the critical temperature probability distribution is presented.
Finally, in \sref{sec:sec5}
we conclude with a relation between the values of the critical exponents in
the pure and in the random systems.

\section{Mean-Field McCoy-Wu model}
\label{sec:sec2}
\begin{figure}
\epsfxsize=9cm
\begin{center}
\mbox{\epsfbox{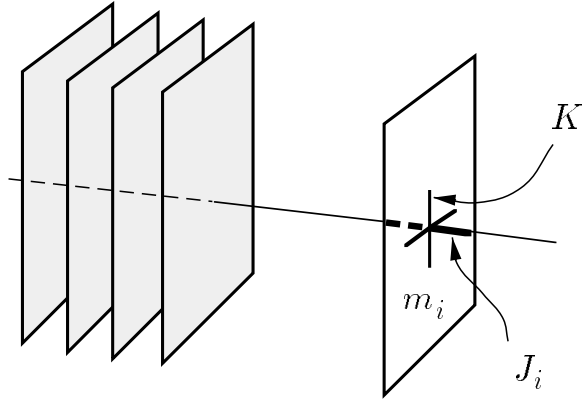}}
\end{center}
\vskip 0mm
\caption{$d$-dimensional layered mean-field model} 
\label{fig1}  
\end{figure}

As mentioned in the Introduction we consider a $d$-dimensional Ising
model, which consists of $(d-1)$-dimensional layers, such that the Hamiltonian
is given by:
\be
H=-\sum_i \sum_j J_i \sigma_{i,j} \sigma_{i+1,j} - K \sum_i \sum_{<j,k>}
\sigma_{i,j} \sigma_{i,k}\;.
\label{hamilton}
\ee
Here $\sigma_{i,j}= \pm 1$ and $i=1,2,\dots,L$ characterises the position of
the layers, whereas $j$ and $k$ give the position of the spin within a layer
and $<j,k>$ are nearest neighbours. We treat the Hamiltonian in
(\ref{hamilton}) in mean-field theory, then the
local
magnetization in the $i$-th layer, $m_i=<\sigma_{i,j}>$ 
(see \fref{fig1}), is subject of
variation, if the $J_i$ couplings are inhomogeneous. According to local
mean-field theory the local magnetization satisfies the following set of
self-consistency equations:
\be
m_i={\rm tanh} \left[{J_{i-1} m_{i-1} + 2(d-1)K m_i + J_i m_{i+1} \over T}
\right]\;,
\label{selfeq}
\ee
for $i=1,2,\dots,L$ and with $m_0=m_{L+1}=0$. 

From here on we use units with
$k_B=1$. The self-consistency equations in (\ref{selfeq}) have to be supplemented
by boundary conditions (b.c.). Here
we apply symmetry breaking b.c., such that the spins in one surface layer ($i=1$)
are free, thus $J_0=0$, whereas in the
other surface layer ($i=L$) they are fixed to the same state, thus $m_L=1$.
The advantage of this type
of b.c. is twofold: 
\begin{description}
\item{i)} one can study both the bulk and surface quantities at the
same time, and 
\item{ii)} one can investigate the profiles also at and above the
critical temperature.
\end{description}

As we already mentioned the $J_i$ exchange couplings are quenched random
variables. It is generally assumed that the average behaviour of the physical
quantities does not depend on the details of the distribution of the couplings.
In the following,  we use the symmetric binary distribution:
\be
\rho(J)={1 \over 2} \delta(J-\lambda) + {1 \over 2} \delta(J-\lambda^{-1})\;,
\label{binary}
\ee
furthermore, to reduce the number of parameters we take $(d-1)K=\lambda$.

In this paper the MFMW model is studied numerically on finite slabs with
relatively large
width $(L \le 1024)$, such that for a given random realization of the
couplings the self-consistency equations in (\ref{selfeq}) are solved
by the Newton-Raphson
method. The resulting magnetization profile is then averaged over several
$(\sim 10^5)$ samples. 

\begin{figure}
\epsfxsize=11cm
\begin{center}
\mbox{\epsfbox{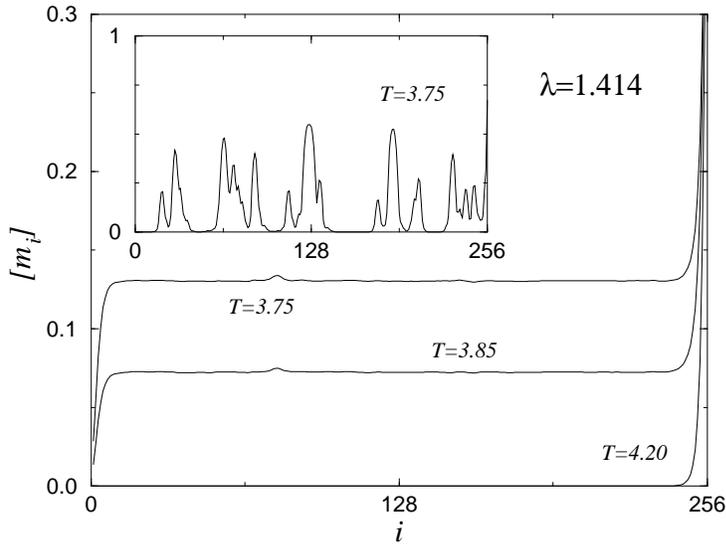}}
\end{center}
\vskip -5mm
\caption{Averaged order-parameter profiles with free-fixed boundary 
conditions
on a finite system of width $L=256$  at different temperatures
below  and around the critical temperature ($T_c^{av}=4.223$),
for $\lambda=1.414$. 
The insert shows a  specific disorder realization below $T_c^{av}$.} 
\label{fig2}  
\end{figure}

According to the numerical results, the MFMW model exhibits two phases which
are separated by a critical point at $T_c^{av}$. Above the critical
temperature, $T>T_c^{av}$, the average bulk magnetization is zero
and the magnetization profile at $i=L$ drops to zero within the range of
the surface correlation length
$\xi_{\perp} \sim |T-T_c^{av}|^{-\nu}$, where $\nu$ denotes the
corresponding critical exponent. Below the critical temperature,
$T<T_c^{av}$, the average magnetization is finite at any site of
the system. As seen in \fref{fig2} the average bulk magnetization $[m_b]_{av}$
corresponds
to the value of $m$ in the plateau of the profile, which is different
from the surface magnetization, and $[m_b]_{av}>[m_1]_{av}>0$.
Again the width of
the two surface regions, both at $i=1$ and $i=L$, are characterized by the corresponding
correlation lengths.

\section{Numerical determination of the critical exponents}\label{sec:sec3}

\begin{figure}
\epsfxsize=11cm
\begin{center}
\mbox{\epsfbox{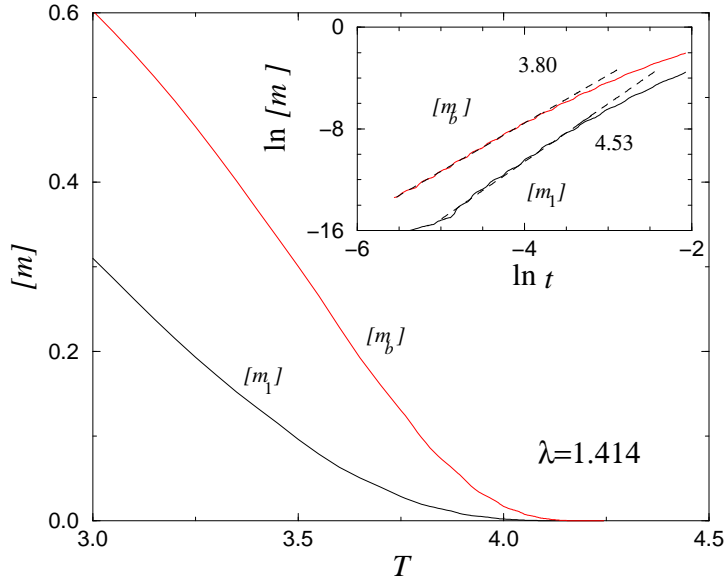}}
\end{center}
\vskip 0mm
\caption{Temperature dependence of the average bulk and local surface 
magnetizations (disorder amplitude $\lambda=1.414$). The 
corresponding log-log plots are shown  in the insert, where the dashed lines
correspond to a linear fit leading to approximate values $\beta\approx 3.80$
and $\beta_1\approx 4.53$.} 
\label{fig3}  
\end{figure}

The temperature dependence of the bulk and surface magnetization is
shown in \fref{fig3}. As seen in the figure both $[m_b]_{av}$ and $[m_1]_{av}$
vanish at the same temperature, thus we have the so-called {\it ordinary
surface transition}~\cite{binder}. The magnetizations close to the critical
point are described by power laws in terms of the 
reduced temperature $t=T_c^{av}-T$
as $[m_b]_{av}(t) \sim t^{\beta}$ and $[m_1]_{av}(t) \sim
t^{\beta_1}$, respectively. Indeed, as seen in the insert in \fref{fig3} the magnetizations versus reduced
temperature in a log-log plot are asymptotically described by straight
lines, the slope of those are given by the corresponding magnetization
exponents.

\begin{figure}
\epsfxsize=11cm
\begin{center}
\mbox{\epsfbox{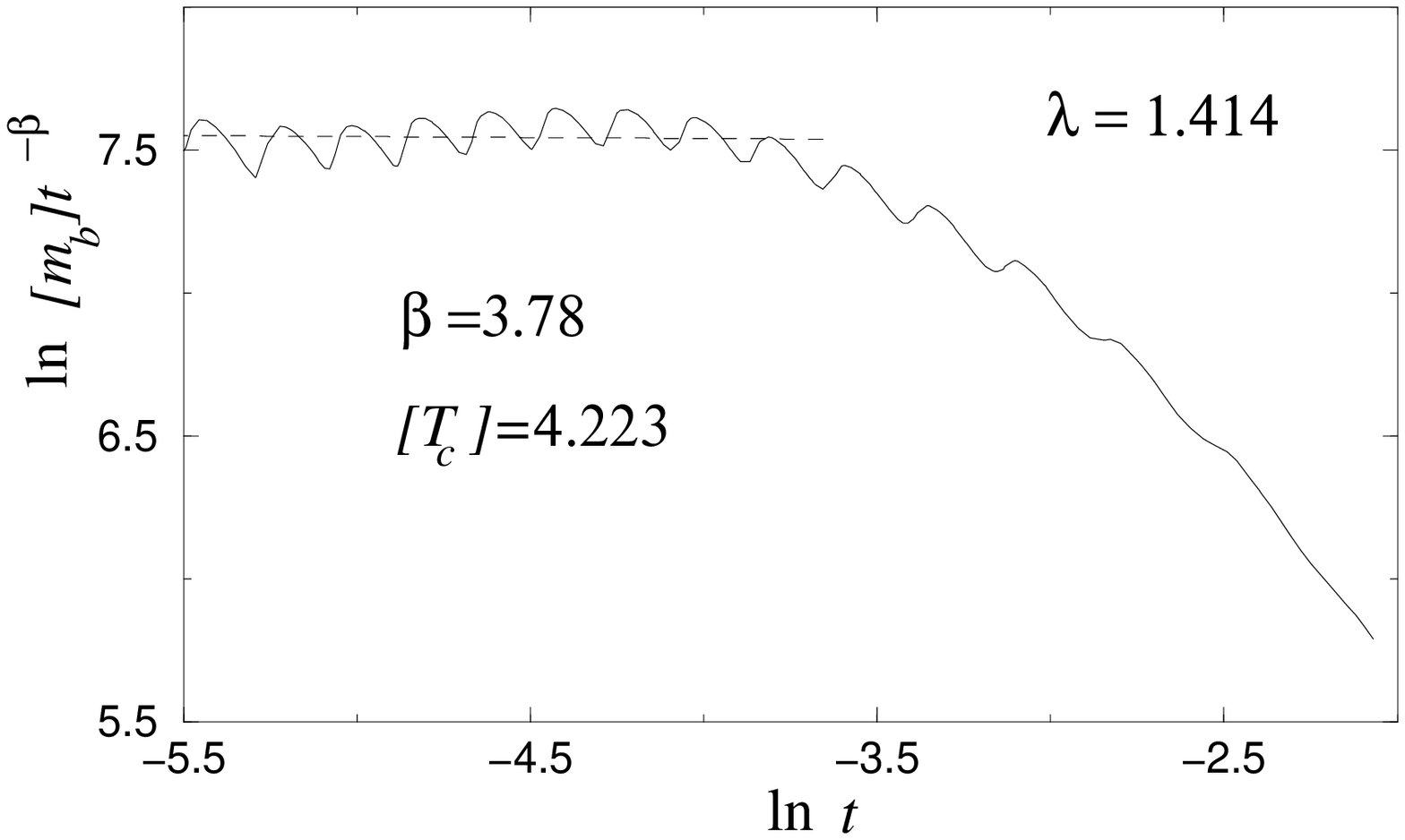}}
\end{center}
\vskip -20mm
\caption{Rescaled average bulk magnetization at $\lambda=1.414$ with
log-periodic oscillations, which are used to obtain refined estimates both
on the critical temperature and on the critical exponent $\beta$.}
\label{fig4}  
\end{figure}

Having a closer look to \fref{fig3} one can notice that the magnetization
close to the critical point exhibits log-periodic oscillations as a
function of $t$~\cite{karevskiturban96}. The origin of these oscillations is the existence of a
finite energy scale in the binary distribution in  \eref{binary},
which is connected to the difference between the
two possible values of the couplings $\lambda$ and
$\lambda^{-1}$~\cite{note}.
We use these log-periodic oscillations to improve our
estimates on the critical temperature and on the critical exponents, at
the same time.
The resulting reduced magnetization $[m_b]_{av} t^{-\beta}$ versus $t$
is presented in \fref{fig4} on a log-log plot, where we have taken  optimized values for
$\beta$ and $T_c^{av}$. In this figure we used the critical temperature to
obtain perfect
oscillations, whereas the correct value of the critical exponent $\beta$
is connected with a constant asymptotic limit of $[m_b]_{av} t^{-\beta}$
as $t \to 0$.

The estimated critical temperatures, together with the bulk and surface
magnetization exponents are given in Table 1 for different values of the
parameter $\lambda$ of the binary distribution. As seen, the critical
exponents do not depend on the strength of  randomness and they agree,
within the error of the estimates, with each other:
\be
\beta=3.6(2)~~,~~~\beta_1=4.2(2)\;.
\label{beta}
\ee
We note that these exponents are unconventionally large, especially if we
compare them with the similar ones of the pure model. A large $\beta$
exponent is connected with a fast variation of the magnetization around the
critical point and the critical region in $t$, where the substantial variation
of $[m]_{av}(t)$  takes place, is then very narrow. Therefore in a numerical
calculation of the critical exponents one should approach closely the
critical point, which in turn will lead to an increase of the error of the
estimation. This fact explains the
not very high accuracy of the numerical values in  \eref{beta}.

\begin{table}
\caption{Numerical values of the critical temperature and the magnetic
exponents for the surface and  bulk magnetizations. }
\footnotesize\rm
\begin{indented}
\item[]\begin{tabular}{@{}llll}
\br
$\lambda$ & $T_c^{av}$ & $\beta$ & $\beta_1$ \\
\mr
 1.414 & 4.223  & 3.78  &4.43 \\
  2. & 4.969  & 3.60  &4.33 \\
 3.162 & 6.908  & 3.51  &4.26 \\
  \br
\end{tabular}
\end{indented}
\label{table1}
\end{table}

The same fact, the relatively large values of the magnetization exponents, have
made it very difficult to obtain a numerical estimate on the correlation length
exponent $\nu$. In principle it can be determined from the decay of the
magnetization profile at the critical point, which, according to the
Fisher-de Gennes scaling theory~\cite{fisherdegennes} asymptotically
behaves as:
\be
[m(l)]_{av} \sim l^{-\beta/\nu}\;,
\label{decay}
\ee
where $l=L-i$.
For the MFMW model, however, due to the large value of $\beta$ the decay
in \eref{decay} is very fast and the profile will become smaller than
the noise before its asymptotic regime is reached. Therefore we were not able to
obtain a sensitive value for $\nu$.

\begin{figure}
\epsfxsize=11cm
\begin{center}
\mbox{\epsfbox{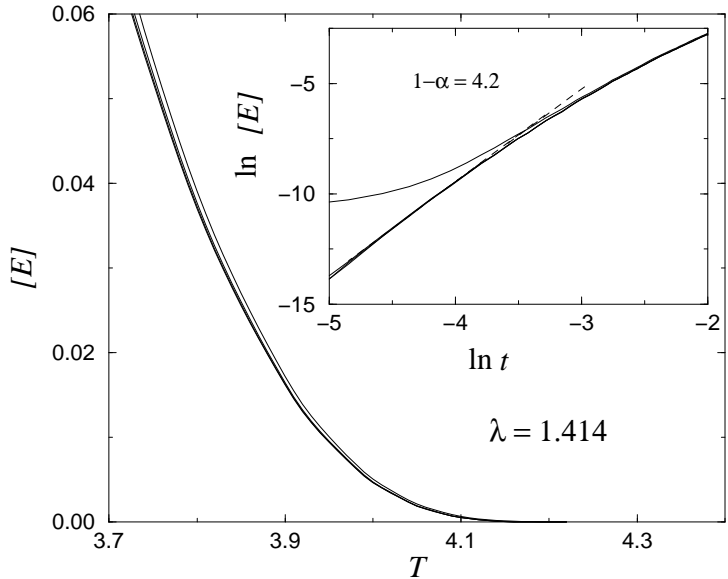}}
\end{center}
\vskip -0mm
\caption{Temperature dependence of the internal energy and corresponding
log-log plot in insert. The different curves correspond
to different chain sizes (from $L=32$ to 256) and the finite-size effects are
quite small.} 
\label{fig5}  
\end{figure}

Next we consider the specific heat of the system, the critical behaviour
of which is deduced from the average internal energy:
\be
[E]_{av}=-\sum_i\left[ J_i m_{i-1} m_i + 2(d-1)K m_i^2 \right]_{av}\;,
\label{energy}
\ee
as $C_v={1 \over N} {\delta [E]_{av} \over \delta T}$. As seen in \fref{fig5}
the specific heat at the critical point has a power law singularity and
the corresponding critical exponent is obtained from the slope of the
curve in a log-log scale as:
\be
\alpha=-3.2(1)\;.
\label{alpha}
\ee
For the specific heat exponent, similarly to the magnetization exponents, we
have made use of the log-periodic nature of the oscillations to increase the
accuracy of the estimates. We note that the specific heat exponent
in (\ref{alpha}) is negative, thus it is decreased from its pure value
$\alpha_p=0$ and consequently, due to randomness
the specific heat has become less singular. The same observation was reported
for
a marginally aperiodic layered Ising model in mean-field theory~\cite{inhomogen1}.

\section{Probability distribution of the critical temperature}\label{sec:sec4}

After having determined the {\it average} values of the physical
quantities, which are accessible in a measurement, we are going
to study their probability distributions. In this respect the distribution
of the samples critical temperature $T_c$ is of primary importance.
For a given random realization of the $J_i$ couplings, the critical
temperature is obtained from \eref{selfeq} in the limit $m_i \to 0$.
Then one proceeds by replacing in the r.h.s. of (\ref{selfeq}) the
${\rm tanh}(x)$ by $x$ and solve the linear eigenvalue problem
\be
\left(
\matrix{
a_T & J_1 &  0  &       & \dots &       &    0  \cr
J_1 & a_T & J_2 &       &       &       &       \cr
 0  & J_2 & a_T & J_3   &       &       &       \cr
    &     & J_3 & a_T   &\ddots &       &  \vdots     \cr
\vdots&     &      &\ddots&\ddots &J_{L-2}&    0   \cr
    &     &      &      &J_{L-2}&  a_T  &J_{L-1}\cr
0   &     &      &  \dots&  0    &J_{L-1}&  a_T  \cr}
\right)
 \left(\matrix{
m_1\cr
m_2\cr
m_3\cr
 \vdots\cr
m_{L-2}\cr
m_{L-1}\cr
m_L\cr}
\right)=0\;,
\label{tc}
\ee
for the critical temperature $T_c$, which is contained in the diagonal
term, since $a_T=2(d-1)K-T_c$.

\begin{figure}
\epsfxsize=13cm
\begin{center}
\mbox{\epsfbox{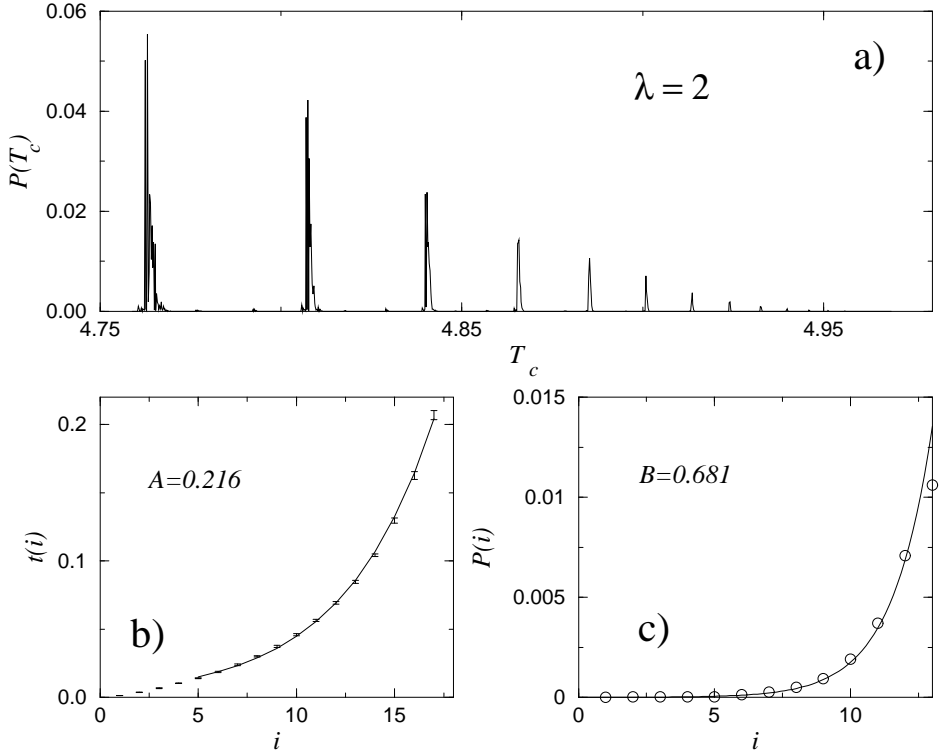}}
\end{center}
\vskip -0mm
\caption{Probability distribution of the critical temperature and its
behaviour: a) Distribution of the samples critical temperatures, b)
Exponential fit of relative critical temperatures $t(i)=T_c^{max}-T_c(i)$,
c) Exponential fit of the corresponding weigth $P(i)$.} 
\label{fig6}  
\end{figure}

The distribution of the samples critical temperature is shown in \fref{fig6}a
for the parameter $\lambda=2$ of the binary distribution \eref{binary},
but similar type of behaviour is found for all other values of $\lambda$. As seen
in \fref{fig6}a the distribution consists of sharp peaks the widths of those
is much smaller than the distance between them. We shall number the peaks by
$i=0,1,\dots$ in descending order from the
maximal one and denote by $T_c(i)$ the characteristic value of the critical
temperature measured at the position of the tip of the peak.
Thus we have $T_c(i=0)=T_c^{max}$ and
$t(i)=T_c^{max}-T_c(i)$ measures the difference from the maximal critical
temperature. First we
note that, within the error of the calculation, the $T_c^{max}$ maximal
critical temperature is equal to the average critical temperature
\be
T_c^{max}=T_c^{av}\;,
\label{tcmax}
\ee
which has been determined before from
the behaviour of the average magnetization and the specific heat. 
We note that $T_c^{max}$ in (10) corresponds to the so called
{\it Griffiths temperature} in random (Ising) spin systems, which is
just the upper border of the Griffiths phase. In our system the observation
in (10), i.e. $T_c^{max}$ and the Griffiths temperature coincides, means
that there is no realization which exhibit finite bulk magnetization above the average
critical temperature $T_c^{av}$. As a consequence the average quantities,
such as the susceptibility, behave analytically above the critical temperature,
thus there are no Griffiths singularities in the system. We note 
that similar observation is found in random systems with long
range interactions, where mean-field theory is exact~\cite{MFSG}.

\begin{figure}
\epsfxsize=10cm
\begin{center}
\mbox{\epsfbox{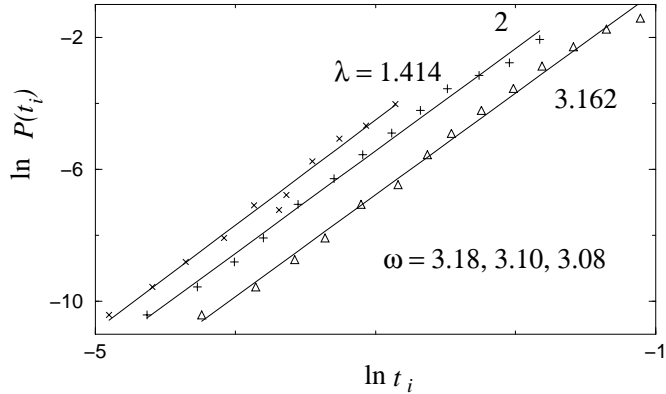}}
\end{center}
\vskip -20mm
\caption{Power-law behaviour of the critical temperature distribution with respect
to the difference from the maximal critical temperature, $t_i$, estimated at the 
successive peaks for
different values of the disorder amplitude: $\lambda=1.414$ ($\times$),
$\lambda=2.$ ($+$), and $\lambda=3.162$ ($\triangle$). The corresponding
values of $\omega$ are given in the figure in the same order.} 
\label{fig7}  
\end{figure}

In the following we study the $t(i)=T_c^{max}-T_c(i)$
relative critical temperatures and the corresponding weigth $P(i)$ as a
function of the
index of the peak, $i$. As seen on \fref{fig6}bc both quantities could be well
fitted by exponential functions~\cite{notesinai}:
\be
t(i) \sim \exp(Ai)~~,~~~P(i) \sim \exp(Bi)\;.
\label{tP}
\ee
The $A$ and $B$ parameters in
\eref{tP} are found approximately independent of the form of the random
distribution of the couplings and they ratio is given by:
\be
\omega={B \over A}=3.1(1)\;.
\label{omega}
\ee
Combining the two relations in \eref{tP} we obtain a power law dependence
of the $P(t_i)=P(i)$
probability distribution:
\be
P(t_i) \sim t_i^{\omega}\;,
\label{pt}
\ee
with $\omega$ given in \eref{omega}. This relation is indeed well
satisfied, as can be seen in \fref{fig7}.

\section{Relation between pure and random system critical exponents}
\label{sec:sec5}

In the following we use the form of the probability distribution in
\eref{pt} to relate the values of the critical exponents of the
pure and the random systems. Generally we consider a physical quantity
$Q(t)$, which behaves in the homogeneous system
\be
Q(t) \sim t^{\epsilon_p}\;,
\label{Qt}
\ee
as a function of the
reduced temperature $t=T_c-T$, for $|t| \ll 1$.
(In mean-field theory for the bulk magnetization
$\epsilon_p=\beta_p=1/2$, for the surface magnetization
$\epsilon_p=\beta_{1_p}=1$ and for the specific heat
$\epsilon_p=-\alpha_p=0$, etc.) We restrict ourselves to
quantities with $\epsilon_p \ge 0$. To calculate the average
behaviour of $Q(t)$ in the random system, we assume that in each random
realization the temperature dependence $Q_i(t)$ is the same as in the pure
system in (\ref{Qt}) with the appropriate critical temperature $T_c(i)$ of the
sample. This relation is then averaged over the samples:
\be
[Q(t)]_{av}=\sum_{t_i>t} P(t_i) Q_i(t) \sim \sum_{t_i>t}
t_i^{\omega}(t_i-t)^{\epsilon_p} \sim t^{\omega+\epsilon_p}\;.
\label{average}
\ee
Thus the critical exponent in the random system, $\epsilon$, is related to
its value in the homogeneous system as:
\be
\epsilon=\epsilon_p+\omega\;.
\label{exprel}
\ee
This relation is indeed satisfied with all the considered physical
quantities in eqs(\ref{beta}) and (\ref{alpha}).

To summarize we have considered a generalized McCoy-Wu model and
studied the critical properties in the mean-field approximation. We have determined different
critical exponents and showed that they do not depend on the actual
form of the coupling distributions. The values of the critical exponents
in the pure and in the random systems are related and the only parameter
which completely characterizes the random critical properties is the
$\omega$ exponent of the probability distribution of the critical
temperatures. We have seen in \eref{tcmax} that the average critical
temperature corresponds to the maximal critical temperature of the
samples. Therefore above the $T_c^{av}$ critical temperature there are no
samples with finite magnetization and hence there are {\it no Griffiths
singularities} in the MFMW model.

The critical properties of the model are deeply connected to the probability
distribution of the samples critical temperature in eqs (\ref{tP}) and
(\ref{pt}). We consider it very probable that these expressions, which
were observed numerically, can be obtained by analytical methods
and perhaps also the $\omega$ exponent in (\ref{omega}) can be determined
exactly.

 \ack 
This work has been supported by the French-Hungarian cooperation program
"Balaton" (Minist\`ere des Affaires Etrang\`eres-O.M.F.B.), by the Hungarian
National Research Fund under grants No OTKA TO12830, OTKA TO23642 and
OTKA TO25139
and by the Ministery of Education under grant No. FKFP 0765/1997. We thank the
CIRIL in Nancy for computational facilities.
We are indebted to H. Rieger and L. Turban for useful discussions.


\section*{References}
\begin{thebibliography}{99}



\bibitem{mccoywu}
        B.M. McCoy and T.T. Wu, 
        Phys. Rev. {\bf 176}, 631 (1968); {\bf 188}, 982(1969);
        B.M. McCoy, Phys. Rev. {\bf 188}, 1014 (1969).

\bibitem{fisher}
        D.S. Fisher, Phys. Rev. Lett. {\bf 69}, 534 (1992); 
        Phys. Rev. B {\bf 51}, 6411 (1995).

         
\bibitem{youngrieger}
        A. P. Young and H. Rieger, 
        Phys. Rev. B {\bf 53}, 8486 (1996).

\bibitem{igloirieger97}
        F. Igl\'oi and H.\ Rieger,
        Phys. Rev. Lett. {\bf 78}, 2473 (1997).

\bibitem{young}
	A. P. Young, Phys. Rev. B {\bf 56}, 11691 (1997).

\bibitem{igloirieger98}
        F. Igl\'oi and H.\ Rieger,
        Phys. Rev. B, in press (1998).

\bibitem{griffiths}
	R.B. Griffiths, Phys. Rev. Lett. {\bf 23}, 17 (1969).

\bibitem{mccoy}
    B. McCoy, Phys. Rev. Lett. {\bf 23}, 383 (1969)

\bibitem{qspglass}
        H. Rieger and A.P. Young, Phys. Rev. Lett. {\bf 72}, 4141 (1994);
        M. Guo, R.N. Bhatt and D.A. Huse, Phys. Rev. Lett. {\bf 72}, 4137 (1994).

\bibitem{MFSG}
	J. Miller and D.A. Huse, Phys. Rev. Lett. {\bf 70}, 3147 (1993);
	N. Read, S. Sachdev and J. Ye, Phys. Rev. B{\bf 52}, 384 (1995).

\bibitem{DFM}
	T. Senthil and S. Sachdev, Phys. Rev. Lett. {\bf 77}, 5292 (1996);
	T. Ikegami, S. Miyashita and H. Rieger, J. Phys. Soc. Jap. (in press).

\bibitem{RFM}
	H. Rieger and N. Kawashima, submitted to Phys. Rev. Lett.;
	C. Pich and A.P. Young, submitted to Phys. Rev. Lett.

\bibitem{spinglass}
	K. Binder and A.P. Young, Rev. Mod. Phys. {\bf 58}, 801 (1986).

\bibitem{inhomogen1}
       P.E. Berche and B. Berche, \JPA\ {\bf 30}, 1347 (1997).
\bibitem{inhomogen2}
       F. Igl\'oi and G. Pal\'agyi, Physica A {\bf 240}, 685 (1997).

\bibitem{binder}
	    K. Binder, in {\it Phase Transitions and Critical Phenomena}, vol 8;
        eds. C. Domb and J.L. Lebowitz, (London: Academic Press), p 1 (1983)

\bibitem{karevskiturban96}
	D. Karevski and L. Turban, \JPA\ {\bf 29}, 3461 (1996).

\bibitem{note}
        Indeed there are no log-periodic oscillations, if the couplings
        follow uniform distribution, where no finite energy scale can be
        defined. 


\bibitem{fisherdegennes}
        M.E. Fisher and P.-G. De Gennes, 
        C.R. Acad. Sc. Paris B {\bf 287}, 207 (1978).

\bibitem{notesinai}
        Somewhat similar, exponential relation is present in the Sinai
        model (Ja. G. Sinai, Theor. Prob. Appl. {\bf 27}, 247 (1982)),
       in a one-dimensional random walk in a random environment, where the
       $\tau$ time- and $L$ length-scales are related as $\tau \sim \exp(AL^{1/2})$.
       We can use analogous language for the MFMW model, if we notice
that the eigenvalue matrix in \eref{tc},
which serves to determine the samples critical temperature, is equivalent
to the transfer matrix of a one-dimensional directed walk,
if a step of the walk on the $i$-th site is weighted by a fugacity $J_i$.
The relevant time-scale of the problem $\tau_w$ is related to the $\Delta$
gap at the top of the spectrum of the
transfer matrix, which is connected to the relative critical temperature
of the MFMW model as $\Delta \sim t_1$.
 
\end{thebibliography}
\end{document}